%% file: orderSchedulingSetups.tex
\documentclass[runningheads]{article}
\usepackage{graphicx}
\usepackage{amsmath}
\usepackage{amsthm}
\usepackage{listings}
\usepackage{graphicx}
\usepackage{enumerate}
\usepackage{mathtools} 
\usepackage{algpseudocode}
\usepackage{todonotes}
\usepackage{hyperref}
\usepackage{xspace}
\usepackage{amsfonts}
\usepackage{subcaption}
\usepackage[capitalise,nameinlink,noabbrev]{cleveref}

\newcommand{\modelMainW}{$1|s_f,~\allowbreak assembly|\allowbreak\sum_{}^{} w_j C_j$\xspace}
\newcommand{\modelOS}{$1|ot\textit{-}s_f,~\allowbreak assembly|\allowbreak\sum_{}^{} w_j C_j$\xspace}
\newcommand{\modelGluedOS}{$1|ot-s_f,~\allowbreak glued~assembly|\allowbreak\sum_{}^{} w_j C_j$\xspace}
\newcommand{\modelDefPrec}{$1|prec|\allowbreak\sum w_j C_j$\xspace}

\newcommand*{\opt}{\ensuremath{C^\textsc{Opt}}\xspace}
\newcommand*{\out}{\ensuremath{C^{\pi_\text{out}}}\xspace}
\newcommand*{\greedy}{\ensuremath{\textsc{Greedy}}\xspace}
\newcommand*{\ggreedy}{\ensuremath{\textsc{Greedy}^+}\xspace}
\newcommand*{\pullfactor}{\ensuremath{\beta}\xspace}

\newtheorem{theorem}{Theorem}
\newtheorem{lemma}{Lemma}
\newtheorem{corollary}{Corollary}

\begin{document}
	\title{Approximating Weighted Completion Time for Order Scheduling with Setup Times\thanks{This work was partially supported by the German Research Foundation (DFG) within the Collaborative Research Centre “On-The-Fly Computing“ under the project number 160364472 --- SFB 901/3.}
		\footnote{This paper was accepted for puplication in the proceedings of the International Conference on Current Trends in Theory and Practice of Computer Science (SOFSEM 2020)}
		}
	%
	%
	\author{Alexander M\"acker \and
		Friedhelm Meyer auf der Heide\and
		Simon Pukrop\\
		Heinz Nixdorf Institute \& Computer Science Department,\\
		Paderborn University, F\"urstenallee 11, 33102 Paderborn\\
		\texttt{\{amaecker, fmadh, simonjp\}@mail.uni-paderborn.de}
	}
	%
	%
	%
	\maketitle              
	\providecommand{\keywords}[1]{\textbf{\textit{Index terms---}} #1}
	\begin{abstract}
		Consider a scheduling problem in which jobs need to be processed on a single machine.
		Each job has a weight and is composed of several operations belonging to different families.
		The machine needs to perform a setup between the processing of operations of different families.
		A job is completed when its latest operation completes and the goal is to minimize the total weighted completion time of all jobs.
		
		We study this problem from the perspective of approximability and provide constant factor approximations as well as an inapproximability result.
		Prior to this work, only the NP-hardness of the unweighted case and the polynomial solvability of a certain special case were known.
		
	\end{abstract}
	%
	%
	\section{Introduction}
	\label{sec:intro}
	Many models for scheduling problems assume jobs to be atomic. 
	There are, however, numerous natural situations where it is more suitable to model jobs as compositions and consider the problem as an \emph{order scheduling} formulation:
	In this case a job is called order and is assumed to be composed of a set of operations, which are requests for products.
	A job is considered to be finished as soon as all its operations are finished and a natural objective is the minimization of the sum of completion times of all jobs.
	
	Another important aspect in such scenarios can be the consideration of setup times that may occur due to the change of tools on a machine, the reconfiguration of hardware, cleaning activities or any other preparation work \cite{ali1,ali2,ali3}.
	We model this aspect by assuming the set of operations to be partitioned into several families.
	The machine needs to perform a setup whenever it switches from processing an operation belonging to one family to an operation of a different family.
	Between operations of the same family, however, no setup is required.
	
	This kind of order scheduling (with setups) has several applications, which have been reported in the literature and we name a few of them here:
	It can model situations in a food manufacturing environment \cite{agreeable}.
	Here, several base ingredients need to be produced on a single machine and then assembled to end products and setup times effectively occur due to cleaning activities between producing different base ingredients.
	Another example \cite{agreeable} are customer orders where each order requests several products, which need to be produced by a single machine, and an order can be shipped to the customer only as soon as all products have been produced.
	Finally, our primary motivation for considering multioperation jobs comes from its applicability within our project on ``On-The-Fly Computing'' \cite{sfb901}.
	Here, the main idea is that IT-services are (automatically) composed of several small, elementary services that together provide the desired functionality.
	Setup times occur due to the reconfiguration of hardware or for the provisioning of data that needs to be available depending on the type of elementary service to be executed.
	
	\subsection{Contribution \& Results}
	We consider the aforementioned problem, which is formally introduced in \cref{sec:model} and which in the survey \cite{orderSurvey} by Leung et al.\ was termed \emph{fully flexible case of order scheduling with arbitrary setup times}, for the case of a single machine.
	Because it is known that the problem is NP-hard as mentioned in \cref{sec:relatedWork} where we summarize relevant related work, we study the problem with respect to its approximability.
	The key ingredient of our approach is based on the following idea.
	We define a simplified variant of the considered problem, in which we only require that, before any operation of a given family is processed, a setup for this family is performed once at \emph{some} (arbitrary) earlier time.
	Solutions to this simplified variant already carry a lot of information for solving the original problem. 
	We show that they can be transformed into $(1+\sqrt{2})$-approximate solutions for our original problem in polynomial time in \cref{sec:oneTime}.
	We then provide an algorithm that solves the simplified variant optimally leading to a $(1+\sqrt{2})$-approximation for the original problem in \cref{sec:constantFamilies}.
	The runtime of the approach, however, is $O(\text{poly}(n) \cdot K!)$, where $K$ denotes the number of families.
	Thus, it is only polynomial for a constant number of families, which turns out to be no coincidence as we also observe that solving the simplified variant optimally for non-constant $K$ is NP-hard.
	We then show how an algorithm by Hall et al.\ \cite{hall} can be combined with our approach from \cref{sec:oneTime} to obtain a runtime of $O(\text{poly}(n, K))$ while worsening the approximation factor to $2(1+\sqrt{2})$ in \cref{sec:variableFamilies}.
	We complement this result by a hardness result for approximations with a factor less than $2$ assuming a certain variant of the Unique Games Conjecture.
	
	Finally, we present some results of a simulation-based evaluation of our approach in \cref{sec:simulation}.
	We show that on randomly created instances our algorithm even performs better than suggested by the approximation factor of $3$ and we show how our approach can be tuned to improve its performance in such settings.
	
	\section{Model}
	\label{sec:model}
	We consider a scheduling problem in which a set $\mathcal{J}=\{j_1, \ldots, j_n\}$ of $n$ jobs needs to be processed by a single machine.
	Each job $j$ has a weight $w(j) \in \mathbb{R}_{\geq0}$ and consists of a set of operations $j = \{o^j_1, o^j_2, \ldots\}$.
	Each operation $o_i^j$ is characterized by a processing time $p(o_i^j) \in \mathbb{R}_{\geq0}$ and belongs to a family $f(o_i^j) \in \mathcal{F} = \{f_1, \ldots, f_{K}\}$.
	If the schedule starts with an operation of family $f$ and whenever the machine switches from processing operations of one family $f'$ to an operation of another family $f$, a setup taking $s(f) \in \mathbb{R}_{\geq0}$ time needs to take place first.
	Given this setting, the goal is to compute a schedule that minimizes the weighted sum of job completion times, where a job is considered to be completed as soon as all its operations are completed.
	More formally, a schedule is implicitly given by a permutation $\pi$ on $\bigcup_{i=1}^n j_i$ and the completion time of an operation $o$ is given by the accumulated setup times and processing times of jobs preceding operation $o$.
	That is, for $\pi=(o_1, o_2, \ldots)$ the completion time of operation $o_i$ is given by $C_{o_i}^\pi = \sum_{k=1}^{i} p(o_k) + \sum_{k=1}^{i} I(f(o_{k-1}),f(o_k))s(f(o_k))$, where $I$ is an indicator being $0$ if and only if its parameters are the same and $1$ otherwise.
	Then, the completion time of a job $j$ is given by $C^\pi_j = \max_{o \in j} C_o^\pi$ and the goal is to minimize the total weighted completion time given by $C^\pi = \sum_{j \in \mathcal{J}}w(j)C_j^\pi$.
	
	Using the classical three-field notation for scheduling problems and following Gerodimos et al. \cite{agreeable}, we denote the problem by \modelMainW. 
	We study this problem in terms of its approximability.
	A polynomial-time algorithm $\mathcal{A}$ has an approximation factor of $\alpha$ if, on any instance, $C^\pi \leq \alpha \cdot \opt$, where $C^\pi$ and \opt denotes the total weighted completion time of $\mathcal{A}$ and an optimal solution, respectively.
	
	\section{Related Work}
	\label{sec:relatedWork}
	The problem \modelMainW, and the more general version with multiple machines, are also known as order scheduling.
	More precisely, it was termed order scheduling in the flexible case with setup times in the survey \cite{orderSurvey} by Leung et al.
	As previously mentioned, it is known that this problem is already NP-hard for the unweighted case and a single machine as proven by Ng et al.\ \cite{hardness}.
	Besides this hardness result, only one single positive result is known. 
	Due to Gerodimos et al.\ \cite{agreeable}, a special case can be solved optimally in time $O(n^{K+1})$.
	This special case requires that the jobs can be renamed so that job $j_{i+1}$ contains, for each operation $o\in j_i$, an operation $o'$ such that $f(o) = f(o')$ and $p(o')>p(o)$.
	A related positive result is due to Divakaran and Saks \cite{saksCompletionTime}.
	They designed a $2$-approximation algorithm for our problem in case all jobs consist of a single operation.
	Monma and Potts worked on algorithms for the same model with a variety of objective functions.
	One result is an optimal algorithm for total weighted completion time with the constraint that the number of families is constant \cite{monma1989complexity}.
	Their approach however relies on the fact that there is a trivial order inside each family, and the problem only arises in interleaving the families.
	Since we are dealing with multi-operation jobs we cannot assume such an order.
	
	Taking a broader perspective of the problem, it can be seen as a generalization of the classical problem of minimizing the total (weighted) completion time when there are no setups and all jobs are atomic (i.e., we only have single-operation jobs).
	It is well-known that sequencing all jobs in the order of non-decreasing processing times (shortest processing time ordering, SPT) minimizes the total completion time on a single machine \cite{lenstra1978complexity}.
	In case jobs have weights and the objective is to minimize the total weighted completion time, a popular result is due to Smith \cite{wspt}.
	He showed that weighted shortest processing time (WSPT), that is, sort the jobs non-decreasingly with respect to their ratio of processing time and weight, is optimal for this objective.
	Besides these two results, the problem has been studied quite a lot and in different variants with respect to the number of machines, potential precedences between jobs, the existence of release times and even more.
	For a single machine it was shown by Lenstra and Kan \cite{lenstra1978complexity} and independently by Lawler \cite{lawler1978sequencing} that adding precedences among jobs to the (unweighted) problem makes it NP-hard.
	In their paper, Hall et al.\ \cite{hall} analyzed algorithms based on different linear programming formulations and obtained constant factor approximations for several variants including the minimization of the total weighted completion time on a single machine with precedences.
	Particularly, they obtained a $2$-approximation for this problem, which we will later make use of for our approximation algorithm for non-constant $K$.
	Actually, the factor $2$ they achieve is essentially optimal, as Bansal and Khot \cite{bansal2009optimal} were able to show that a $(2-\varepsilon)$-approximation is impossible for any $\varepsilon > 0$ assuming a stronger version of the Unique Games Conjecture.
	
	More loosely related is a model due to Correa et al.\ \cite{splittingVsSetups} in which jobs can be split into arbitrary parts (that can be processed in parallel) and where each part requires a setup time to start working on it.
	They proposed a constant factor approximation for weighted total completion time on parallel machines.
	Recently, some approximation results for the minimization of the makespan for single operation jobs have been achieved for different machine environments with setup times \cite{correa15,martenConfigIP,kiel}.
	Finally, scheduling with setup times in general is a large field of research, primarily with respect to heuristics and exact algorithms, and the interested reader is referred to the three exhaustive surveys due to Allahverdi et al. \cite{ali1,ali2,ali3}.
	
	\section{The One-Time Setup Problem}
	\label{sec:oneTime}
	In this section, we introduce a relaxation of \modelMainW and show how solutions to this relaxation can be transformed into solutions to the original problem by losing a small constant factor.
	The \emph{one-time setup} problem (~\modelOS) is a relaxation of \modelMainW in which setups are not required on each change to operations of a different family.
	Instead we only require that, for any family $f$, a setup for $f$ is performed once at some time before any operation belonging to $f$ is processed.
	Formally, we introduce a new (setup) operation $o^s_f$ for each family $f$ with $p(o^s_f) = s(f)$, $w(o^s_f) = 0$ and a precedence relation between $o^s_f$ and each operation belonging to $f$ that ensures that $o^s_f$ is processed before the respective operations.
	A schedule $\pi$ is then implicitly given by a permutation on all operations (those belonging to jobs as well as those representing setups). 
	We only consider those permutations, which adhere to the precedence constraints.
	The completion time of an operation $o_i$ under schedule $\pi=(o_1, o_2, \ldots)$ is given by
	$C_{o_i}^\pi = \sum_{k=1}^{i} p(o_k)$.
	The remaining definitions such as the completion time of a job and total weighted completion time remain unchanged.
	Note that this problem is indeed a relaxation of our original problem in the sense that the total weighted completion time cannot increase when only requiring one-time setups.
	
	Before we turn to our approach to transform solutions to \modelOS into feasible solutions for \modelMainW, we first give a simple observation. 
	It shows that we can, intuitively speaking, glue all operations of a job together and focus on determining the order of such glued jobs.
	Formally, given a schedule $\pi$, a job is \emph{glued} if all of $j$'s operations are processed consecutively without other operations in between. 
	We have the following lemma.
	
	\begin{lemma} \label{lem:glued}
		Every schedule can be transformed into one in which all jobs are glued without increasing the total weighted completion time.
	\end{lemma}
	\begin{proof}
		Consider some schedule $\pi$ with total weighted completion time \opt.
		Without loss of generality assume the jobs to be finished in the order $j_1, j_2, \ldots, j_n$.
		Consider job $j_n$ and let $o \in j_n$ be the operation of $j_n$ processed last.
		We move all operations of $j_n \setminus \{o\}$ so that they are processed consecutively in a block and directly before $o$.
		This does not change the completion time of $j_n$ and does not increase the completion time of any other job.
		Also, because we only move operations to later points in time, all precedence constraints are still satisfied.
		We repeat this process for each job in the schedule in the order $j_{n-1}, j_{n-2}, \ldots, j_1$.
		Thereby, we obtain a schedule with completion time \opt in which all jobs are glued.
	\end{proof}
	
	Due to the previous result, we assume in the rest of the paper that each job $j$ in an instance of \modelOS only consists of a single operation $o^j$.
	This operation has a processing time $p(o^j) = \sum_{o \in j} p(o) \eqqcolon p(j)$ and the precedence relation is extended so that each setup operation with a precedence to some $o \in j$ now has a precedence to $o^j$.
	
	\subsection{Transforming One-Time Setup Solutions}
	\label{sub:transform}
	In this section, we present our algorithm \textsc{Transform} to transform a solution $\pi$ for the one-time setup problem \modelOS into a solution $\pi_\text{out}$ for our original problem \modelMainW.
	Initially, $\pi_\text{out}$ is the sequence of operations as implied by the solution $\pi$ after splitting the glued jobs into its original operations again and not including setup operations.
	A \emph{batch} is a (maximal) subsequence of consecutive (non-setup) operations of the same family. 
	Intuitively, the schedule implied by $\pi_\text{out}$ probably already has a useful order for the jobs but is missing a good batching of operations of the same family.
	This would lead to way too many setups to obtain a good schedule.
	Therefore, the main idea of \textsc{Transform} is to keep the general ordering of the schedule but to make sure that each setup is ``worth it'', i.e., that each batch is sufficiently large to justify a setup.
	We will achieve that by filling up each batch with operations of the same family scheduled later until the next operation would increase the length of the batch too much.
	More precisely, let $B_1, B_2, \ldots$ be the batches in $\pi_\text{out}$ (in this order).
	We iterate over the batches $B_i$ in the order of increasing $i$ and for each batch $B_i$ of family $f$ do the following: 
	Move as many operations of family $f$ from the closest batches $B_{i'}, i' > i$, to $B_i$ as possible while ensuring that $p(B_i) < \pullfactor \cdot s(f)$, where \pullfactor (we call it the \emph{pull factor}) is some fixed constant and $p(B_i) \coloneqq \sum_{j \in B_i} p(j)$.
	If a batch gets empty before being considered, it is removed from $\pi_\text{out}$ (and hence, \emph{not} considered in later iterations).
	We show the following theorem on the quality of \textsc{Transform}.
	
	\begin{theorem} \label{transform3}
		If $C^\pi \leq c \cdot \opt$, then $\out \leq (1+\pullfactor) \cdot c \cdot \opt$, for any $\pullfactor \geq \sqrt{2}$.
	\end{theorem}
	\begin{proof}
		We only need to show that $C^{\pi_\text{out}} \leq (1+\pullfactor) \cdot C^\pi$.
		For the analysis we will compare the completion time of each operation $o$ in $\pi$ to the one in $\pi_\text{out}$ (in their respective cost model).
		We denote by $\pi(\dots o)$ the schedule $\pi$ up to and including operation $o$ and by $f \in \pi(\dots o)$ that some operation in $\pi(\dots o)$ is of family $f$.
		We have 
		\begin{align*}
		C_o^\pi = \sum_{f\in \mathcal{F} | f \in \pi(\dots o)}^{} \underbrace{ \left( p(o_f^s) + \sum_{o' \in \bigcup_{j\in \mathcal{J}} j | o' \in  \pi(\dots o) \land f(o')=f}^{}p(o') \right)}_{\eqqcolon(C_o^\pi)_f} .
		\end{align*}
		We will now analyze the contribution $(C_o^{\pi_\text{out}})_f$ of some family $f$ to the completion time of $o$ in $\pi_\text{out}$.
		We have
		\begin{align*}
		(C_o^{\pi_\text{out}})_f &\leq (C_o^\pi)_f \underbrace{- s(f)}_{\text{removed }o^s_f} + \underbrace{\pullfactor\cdot s(f)}_{\text{added operations}} + \underbrace{\left(\left \lceil{\frac{(C_o^\pi)_f-s(f)}{\frac{\pullfactor}{2}s(f)}}\right \rceil \cdot s(f)\right)}_{\text{cost of setups}}
		\end{align*}
		due to the following reasoning.
		The first three summands describe the contribution of class $f$'s jobs to the completion time of $o$ in $\pi_\text{out}$.
		Compared to $(C_o^\pi)_f$, we move operations of length at most $\pullfactor\cdot s(f)$ belonging to family $f$ in front of $o$ (recall that empty batches are removed in the process of \textsc{Transform}; only the last batch of some family $f$ before $o$ pulls operations from behind $o$ in front of $o$) and we do not consider the one-time setup operation.
		The last summand represents the contribution due to setups for family $f$.
		We need to do at most $\left \lceil{\frac{(C_o^\pi)_f - s(f)}{\frac{\pullfactor}{2}s(f)}}\right \rceil$ many setups for operations of family $f$ that contribute to the completion time of $o$ in $\pi_\text{out}$.
		This is true because of the following reasoning. 
		From our construction we know that for two batches of the same family, with no other batches of the same family in between, the processing time of those batches combined has to be at least $\pullfactor\cdot s(f)$, otherwise they would have been combined.
		If there is an odd number of batches we cannot say anything about the last batch, except that it has a nonzero processing time.
		This factor is captured by the rounding.
		Therefore we obtain
		\begin{align*}
		(C_o^{\pi_\text{out}})_f &\leq
		(C_o^\pi)_f + (\pullfactor-1)s(f) + \left(\left \lceil{\frac{(C_o^\pi)_f - s(f)}{\frac{\pullfactor}{2}s(f)}}\right \rceil \cdot s(f)\right)\\
		&\leq (C_o^\pi)_f + \pullfactor\cdot s(f) + \left( \frac{(C_o^\pi)_f}{\frac{\pullfactor}{2}s(f)} \cdot s(f)\right) - \frac{2}{\pullfactor}s(f) \\
		&\leq (C_o^\pi)_f + (\pullfactor-\frac{2}{\pullfactor})\cdot s(f) + \frac{2(C_o^\pi)_f}{\pullfactor} \\
		&\leq (1+\frac{2}{\pullfactor})(C_o^\pi)_f + (\pullfactor-\frac{2}{\pullfactor})\cdot s(f)
		\stackrel{\pullfactor \geq \sqrt{2}}{<} (1+\pullfactor) (C_o^\pi)_f,
		\end{align*}
		where the last inequality holds because a family $f$ can only contribute to the completion time of $o$ in $\pi_\text{out}$ if it contributed to the completion of $o$ in $\pi$ and in this case $(C^\pi_o)_f \geq s(f)$ by definition.
		(If $o$ itself got moved to the front there might be a family that contributed to $C_o^{\pi}$ but does not to $C_o^{\pi_\text{out}}$).
		Since each operation's completion time in $\pi_\text{out}$ is at most (1+\pullfactor) times as big as in $\pi$, we know that for each job $j \in \mathcal{J}$, $C_j^{\pi_\text{out}} \leq (1+\pullfactor)\cdot C_j^{\pi}$.
	\end{proof}
	
	Actually, one can show that there are instances in which $C^{\pi_\text{out}} \geq (1+\pullfactor) \cdot C^\pi$ and therefore, that the analysis of \textsc{Transform} is indeed tight (cf.\ \cref{sec:tightness}).
	However, it is also worth mentioning that these instances are rather ``artificial'' as the jobs' processing times (and even their sum) are negligible while setup operations essentially dominate the completion times.
	In less nastily constructed instances, we would expect that even for moderate values $\pullfactor > \sqrt{2}$,  $(1+\frac{2}{\pullfactor})(C_o^\pi)_f$ significantly dominates $(\pullfactor-\frac{2}{\pullfactor})\cdot s(f)$ for most of the operations $o$ as $(C_o^\pi)_f$ grows the later $o$ is scheduled while $s(f)$ stays constant.
	This would then lead to  $C^{\pi_\text{out}} \approx 1+\frac{2}{\pullfactor}$.
	This observation is also discussed and supported by our simulations (cf.\ \cref{sec:simulation}).
	
	\subsubsection{Tightness of the Analysis of \textsc{Transform}}
	\label{sec:tightness}
	In \cref{sub:transform} we mentioned that the loss of factor $(1+\pullfactor)$ that we analyzed our \textsc{Transform} algorithm to achieve is tight, for $\pullfactor \geq \sqrt{2}$. 
	We prove that statement formally by providing an instance achieving this factor.
	
	\begin{lemma}
		The analysis of \textsc{Transform} is tight.
	\end{lemma}
	
	\begin{proof}
		\begin{figure}
			\centering
			\def\svgwidth{0.9\linewidth}
			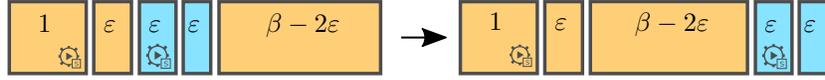
			\caption[]%
			{\small Tightness example for \textsc{Transform} (before and after \textsc{Transform}). Rectangles represent operations and values their respective processing times. Colors represent families and gears the respective setup operations.} 
			\label{fig:tight}
		\end{figure}
		As we can see in \Cref{fig:tight}, for an infinitely small $\varepsilon$, the analysis of $(C_o^{\pi_\text{out}})_f < (1+\pullfactor)\cdot (C_o^\pi)_f$ and $C_o^{\pi_\text{out}} < (1+\pullfactor)\cdot C_o^\pi$ is tight for the blue operation.
		Imagine that both orange operations, as well as the blue operation belong to single operation jobs.
		We replace the singular blue operation job with $m$ many of those operations, each with processing time $\varepsilon$.
		We again compare the total completion time of the schedule, both before and after \textsc{Transform} (in the respective cost models).
		We get $C^\pi \approx 1 + m\cdot1 + (1+\pullfactor)$ while $C^{\pi_\text{out}} \approx 1 + (1+\pullfactor) + m\cdot (1+\pullfactor)$.
		For $m$ to infinity and $\varepsilon$ to zero we get that our analysis of $C^{\pi_\text{out}}\leq (1+\pullfactor)\cdot C^\pi$ is tight.
		Note that we use $\varepsilon$ instead of zero in this example to show that the analysis is tight even when we do not allow processing times of zero.
	\end{proof}
	
	\section{Approximations for Constant Number of Families}
	\label{sec:constantFamilies}
	In this section, we study approximations for the problem \modelMainW and a fixed number of families $K$.
	The general idea is to first solve the \modelOS problem optimally and then to use the \textsc{Transform} algorithm as described in the previous section, leading to $(1+\sqrt{2})$-approximate solutions for instances of  \modelMainW.
	To solve the problem \modelOS optimally, we describe a two-step algorithm and two possible approaches for its second step. 
	The first one is a direct application of a known approach by Lawler \cite{lawler1978sequencing}.
	We also propose a new, alternative approach, which is much simpler as it is specifically tailored to our problem.
	
	To solve \modelOS optimally, in the first step, we exhaustively enumerate all possible permutations of setups.
	In the second step, we then find, for each permutation, the optimal schedule under the assumption that the order of setup operations is fixed according to the permutation.
	After we have performed both of these steps, we can simply select the best result, which is the optimal solution to the \modelOS problem.
	
	\subsection{Series Parallel Digraph and Lawler's Algorithm}
	\label{sub:seriesParallelByLawler}
	Lawler \cite{lawler1978sequencing} proposed an algorithm that optimally solves \modelDefPrec in polynomial time under the condition that the precedences can be described by a series parallel digraph.
	To solve \modelOS, the general idea is to modify the precedence graph of a given \modelOS instance so that it becomes series parallel and then apply Lawler's algorithm.
	We create a series parallel digraph that represents both the jobs reliance on setups as well as the predetermined order of setup operations as follows.
	Given a permutation $\tau = (o_{f_1'}^s, o_{f_2'}^s,\dots)$ of setup operations, we create a precedence chain of nodes $o_{f_1'}^s \rightarrow o_{f_2'}^s \rightarrow \dots$.
	Then for each operation $o^j$, we add an edge from $o_{f_i'}^s$ to $o^j$ such that $i$ is the smallest value for which all operations in $j$ belong to a family in $\{{f_1'},{f_2'},\dots,{f_i'}\}$.
	Intuitively, since we have fixed the order of setups for each operation we can easily see which setup operation is the last one that is necessary to process the operation.
	We do not care about the other precedences because they became redundant after fixing the setup order.
	
	Having done this we have a \modelDefPrec problem with a series parallel digraph that is equivalent to the \modelOS problem.
	At this point we can use the result by Lawler \cite{lawler1978sequencing} to solve this in polynomial time.
	
	\subsection{Simple Local Search Algorithm}
	\label{sub:localSearch}
	
	In this section, we propose a simple algorithm to solve \modelOS optimally in polynomial time given that $K$ is fixed.
	Since our algorithm is tailored to this specific problem, it is a lot simpler and works with less overhead which justifies introducing it here alongside the aforementioned solution.
	
	We first show that we can assume optimal schedules to fulfill a natural generalization of the weighted SPT-order to the setting with setup times.
	We define this notion for our original problem as follows:
	A schedule $\pi$ is in \emph{generalized weighted SPT-order} if the following is true: 
	For every $j_i, j_k$ with $\frac{p(j_i)}{w(j_i)} < \frac{p(j_k)}{w(j_k)}$, $j_i$ is scheduled before $j_k$ or $j_k$ is scheduled at a position where $j_i$ cannot be scheduled (because precedences would be violated).
	If $\frac{p(j_i)}{w(j_i)} = \frac{p(j_k)}{w(j_k)}$, $j_\ominus$ is scheduled before $j_\oplus$ or $j_\oplus$ is scheduled at a position where $j_\ominus$ cannot be scheduled, where $\ominus= \min \{i,k\}$ and $\oplus=\max\{i,k\}$.
	
	\begin{lemma} \label{lem:tranGWSPT}
		Any schedule $\pi$ with total weighted completion time \opt can be transformed into one in generalized weighted SPT-order without increasing the total weighted completion time.
	\end{lemma}
	\begin{proof}
		If there are two jobs $j_i$, $j_k$ with $\frac{p(j_i)}{w(j_i)} < \frac{p(j_k)}{w(j_k)}$ that do not fulfill the desired property, $\pi$ cannot be optimal due to the following reasoning.
		Let $J$ be the set of operations scheduled after $j_i$ and before $j_k$.
		Let $p(J) = \sum_{o \in J} p(o)$ and $w(J) = \sum_{o \in J} w(o)$ denote the processing time of all jobs and setups and all weights in $J$, respectively.
		We show that moving $j_k$ directly behind $j_i$ ($\text{move}_1$) or moving $j_i$ directly before $j_k$ ($\text{move}_2$) reduces the total weighted completion time of $\pi$.
		
		The change of the total weighted completion time due to $\text{move}_1$ is given by $\Delta_1 = -w(J)p(j_k) + p(J)w(j_k) - w(j_i)p(j_k) + w(j_k)p(j_i)$.
		If $\Delta_1 < 0$, $\text{move}_1$ decreases the total weighted completion time and we are done.
		
		Otherwise, if $\Delta_1 \geq 0$, we show that $\text{move}_2$ leads to a decrease.
		Since $\frac{p(j_i)}{w(j_i)} < \frac{p(j_k)}{w(j_k)}$ we know that $- w(j_i)p(j_k) + w(j_k)p(j_i) < 0$.
		Therefore, $-w(J)p(j_k) + p(J)w(j_k) > 0$.
		The change in total weighted completion time $\Delta_2$ of $\text{move}_2$ is given by $\Delta_2 = -(-w(J)p(j_i) + p(J)w(j_i)) - w(j_i)p(j_k) + w(j_k)p(j_i)$.
		Since $\frac{p(j_i)}{w(j_i)} < \frac{p(j_k)}{w(j_k)}$ there exist $x,y \in \mathbb{R}^+$ with $p(j_i) = x\cdot p(j_k)$ and $w(j_i) = x\cdot w(j_k) + y$.
		Plugging those in we get
		\begin{align*}
		\Delta_2 &= -(-w(J)p(j_i) + p(J)w(j_i)) - w(j_i)p(j_k) + w(j_k)p(j_i) \\
		&= -(-w(J)xp(j_k) + p(J)xw(j_k) + p(J)y) - w(j_i)p(j_k) + w(j_k)p(j_i) \\
		&= \underbrace{-x\cdot(-w(J)p(j_k) + p(J)w(j_k))}_{<0} \underbrace{- p(J)y}_{<0} \underbrace{- w(j_i)p(j_k) + w(j_k)p(j_i)}_{<0} < 0.
		\end{align*}
		Therefore, in both cases we get a contradiction to the optimality of $\pi$ and hence, no such jobs $j_i$ and $j_k$ can exist.
		
		It remains to argue about pairs of jobs $j_i$ and $j_k$ such that $\frac{p(j_i)}{w(j_i)} = \frac{p(j_k)}{w(j_k)}$.
		For pairs of jobs $j_i$ and $j_k$ such that $\frac{p(j_i)}{w(j_i)} = \frac{p(j_k)}{w(j_k)}$ we use an argument analogous to the one in the proof above, $\text{move}_1$ or $\text{move}_2$ does not increase the total weighted completion time.
		Additionally, it establishes the desired property between $j_i$ and $j_k$ and one can easily verify that such a move cannot lead to a new violation of the property for any other pair of jobs.
		Therefore, the number of pairs of jobs violating the desired property strictly decreases.
		Repeated application of this process leads to a schedule with the desired properties.
	\end{proof}
	
	Due to the previous lemma, we will restrict ourselves to schedules that are in generalized weighted SPT-order.
	We call the (possibly empty) sequence of jobs between two consecutive setup operations in a schedule a \emph{block}.
	We therefore particularly require that in any schedule we consider, the jobs within a block are ordered according to the weighted SPT-order. 
	
	We execute a local search algorithm started on the initial schedule $\pi^\text{init}_\tau$ given by the input setup operation order $\tau$ followed by all jobs in weighted SPT-order (ties are broken in favor of jobs with lower index).
	An optimal schedule is then computed by iteratively improving this schedule by a local search algorithm.
	Given a schedule $\pi$, a \emph{move} of job $j$ is given by the block into which $j$ is placed subject to the constraint that the resulting schedule remains feasible.
	Note that due to our assumption that we only consider schedules in generalized weighted SPT-order, a schedule $\pi$ and a move of a job $j$ uniquely determine a new feasible schedule.
	A move of job $j$ is called a \greedy move if it improves the total weighted completion time and among all moves of $j$, no other move leads to a larger improvement.
	Among all greedy moves for job $j$ we call the one that places $j$ closest to the beginning of $\pi$ \ggreedy move.
	Our local search algorithm iteratively applies, in weighted SPT-order, one single \ggreedy move for each job.
	For ease of presentation, we assume in the following that we have guessed the permutation $\tau$ of setup operations correctly and that in the following the initial schedule in all considerations is always assumed to be $\pi^\text{init}_\tau$.
	
	\begin{lemma}
		\label{le:movesToOpt}
		Each schedule $\pi$ in generalized weighted SPT-order can be reached by applying, in weighted SPT-order, a single move for each job.
		Additionally, each intermediate schedule is in generalized weighted SPT-order.
	\end{lemma}
	\begin{proof}
		Consider the initial schedule $\pi^\text{init}_\tau$ and let $j_1, j_2, \ldots$ be the jobs in weighted SPT-order.
		Now move $\sigma_i$ for job $j_i$ is performed after the moves for $j_{i'}$, $i'<i$ have been performed and it moves $j_i$ to the respective position (i.e., block) to which it belongs in $\pi$.
		Note that after any move $\sigma_i$, the current schedule is in generalized weighted SPT-order since the jobs $j_1, \ldots, j_i$ form a subschedule of $\pi$, the jobs $j_{i+1}, j_{i+2}, \ldots$ form a subschedule of $\pi^\text{init}_\tau$, and $\frac{p(j_k)}{w(j_k)} \leq \frac{p(j_{k'})}{w(j_{k'})}$ for all $k \leq i$ and $k' > i$.
	\end{proof}
	
	Due to the previous lemma, from now on we assume the following.
	A sequence $\langle\sigma_1, \ldots, \sigma_i \rangle$ of moves defines the schedule obtained by applying the moves $\sigma_1, \ldots, \sigma_i$ (in this order) to the respective first $i$ jobs in weighted SPT-order to the initial schedule $\pi^\text{init}_\tau$.
	Our next step is to show that an optimal schedule can be found by \greedy moves.
	
	\begin{lemma}\label{le:greedymoves}
		Suppose there is a sequence $\langle\sigma_1, \ldots, \sigma_n\rangle$ of moves such that the resulting schedule $\pi$ has total weighted completion time \opt.
		Then all moves are \greedy moves.
	\end{lemma}
	\begin{proof}
		Suppose to the contrary that the total weighted completion time is \opt but there is a move among $\sigma_1, \ldots, \sigma_n$ that is not a \greedy one.
		Let $\sigma_i$ be the last move not being a \greedy one and let $B_\text{opt}$ be the block to which $j_i$ is moved by $\sigma_i$.
		Consider all blocks that can be the destination of a \greedy move of $j_i$ in $\langle\sigma_1, \ldots, \sigma_{i-1}\rangle$.
		Among them let $B_\text{greedy}$ be the one closest to $B_\text{opt}$ if all are behind $B_\text{opt}$ and otherwise let $B_\text{greedy}$ be the last one in front of $B_\text{opt}$.
		Let the move $\sigma'_i$ be the move of $j_i$ to $B_\text{greedy}$.
		Observe that moving $j_i$ to $B_\text{opt}$ or any block between $B_\text{opt}$ and $B_\text{greedy}$ is not a \greedy move by the definition of $B_\text{greedy}$ and the fact that $\sigma_i$ is not a \greedy move.
		Therefore, the total weighted completion time of the schedule $\langle\sigma_1, \ldots, \sigma_i\rangle$ is larger than the one of the schedule $\langle\sigma_1, \ldots, \sigma'_i\rangle$.
		We show that also the total weighted completion time of the schedule $\langle\sigma_1, \ldots, \sigma'_i, \ldots, \sigma_n\rangle$ is smaller than the one of $\langle\sigma_1, \ldots, \sigma_i, \ldots, \sigma_n\rangle = \pi$, which is a contradiction.
		To this end, we distinguish two cases depending on the position of $B_\text{opt}$ compared to $B_\text{greedy}$.
		
		We start with the case that $B_\text{opt}$ is in front of $B_\text{greedy}$.
		Let $J^\ell$ be the set of operations in $\langle\sigma_1, \ldots, \sigma'_i\rangle$ processed before $j_i$ and after the $\ell$-th job after the first job of block $B_\text{opt}$.
		We then deduce by the above observations that for every $\ell \geq 0$ with $J^\ell \neq \emptyset$ it holds 
		\begin{equation}
		\sum_{o \in J^\ell} w(o) p(j_i) > w(j_i) \cdot \sum_{o \in J^\ell} p(o). \label{in:improve}
		\end{equation}
		We claim that each job $j_{i'}$ with $i' > i$ is by $\sigma_{i'}$ moved so that it is within or behind $B_\text{greedy}$ or in front of $B_\text{opt}$.
		Assuming the claim to be true, this concludes the proof of the first case as the improvement due to moves $\sigma_{i+1}, \ldots, \sigma_n$ is independent of whether applied to $\langle\sigma_1, \ldots, \sigma_i\rangle$ or $\langle\sigma_1, \ldots, \sigma'_i\rangle$ and hence, $\langle\sigma_1, \ldots, \sigma'_i, \ldots, \sigma_n\rangle$ has a smaller total weighted completion time than $\pi$, contradicting its optimality.
		It remains to prove the claim. 
		Suppose to the contrary that the claim is not true due to a job $j_{i'}$ (if there are several ones take the first one).
		Let $j$ be the last job processed before $j_{i}$ in $\langle\sigma_1, \ldots, \sigma_{i'}\rangle$.
		Let $J'$ be the set of operations in $\langle\sigma_1, \ldots, \sigma_{i'}\rangle$ processed after $j_{i'}$ and not later than $j$.
		Scheduling $j_{i'}$ in $B_\text{greedy}$ instead would increase the total weighted completion time by
		\[
		w(j_{i'}) \cdot \sum_{o \in J'} p(o) - \sum_{o \in J'} w(o) p(j_{i'}) < 0,
		\]
		where the last inequality follows from \cref{in:improve} together with the fact that $J = J^\ell$ for some $\ell$ and the fact that $\frac{p(j_i)}{w(j_i)} \leq \frac{p(j_{i'})}{w(j_{i'})}$.
		Therefore, $\sigma_{i'}$ is not a \greedy move, which contradicts the assumption that all moves after $\sigma_i$ are \greedy moves.
		
		In case $B_\text{opt}$ is behind $B_\text{greedy}$ we can argue as follows.
		Because $\pi$ is in generalized weighted SPT-order, any job $j_{i'}$ with $i' >i$ can only be placed between $B_\text{greedy}$ and $B_\text{opt}$ if $j_i$ cannot be placed there.
		This, however, is not true due to the definition of $B_\text{greedy}$.
		Therefore, by similar arguments as in the previous case, also $\langle\sigma_1, \ldots, \sigma'_i, \ldots, \sigma_n\rangle$ has a smaller total weighted completion time than $\pi$, which contradicts its optimality.
	\end{proof}
	
	The next corollary follows by the previous three lemmas.
	\begin{corollary}
		There is an optimal schedule that can be reached by applying, in weighted SPT-order, a single \greedy move per job.
	\end{corollary}
	%
	
	Using similar arguments as in the proof of the previous lemma, we can finally show that our tie breaker (by which \greedy and \ggreedy moves differ) does not do any harm when searching for an optimal solution.
	
	\begin{lemma}\label{le:greedymovesToOpt}
		Applying, in weighted SPT-order, a \ggreedy move for each job, leads to an optimal schedule.
	\end{lemma}
	\begin{proof}
		We know by the previous result that there are \greedy moves $\sigma_1, \ldots, \sigma_n$ such that $\langle\sigma_1, \ldots, \sigma_n\rangle \eqqcolon \pi $ is an optimal solution, which is in generalized weighted SPT-order.
		
		Assume $\sigma_1, \ldots, \sigma_{i-1}$ are all \ggreedy moves.
		Let $\sigma^+_i$ be the \ggreedy move for $j_i$ given $\langle\sigma_1, \ldots, \sigma_{i-1}\rangle$.
		Obviously, $\langle\sigma_1, \ldots, \sigma_{i-1}, \sigma_i \rangle$ and $\langle\sigma_1, \ldots, \sigma_{i-1}, \sigma^+_i \rangle$ have the same total weighted completion time.
		We claim that the improvement of each $\sigma_{i'}$ with $i' >i$ is the same independent of whether it is applied to $\langle\sigma_1, \ldots, \sigma_i, \ldots \sigma_{i'-1}\rangle$ or $\langle\sigma_1, \ldots, \sigma^+_i, \ldots \sigma_{i'-1}\rangle$.
		The claim together with the previous lemma leads to the fact that $\sigma_1, \ldots, \sigma^+_i, \ldots \sigma_{n}$ are all \greedy moves.
		Consequently, $\langle\sigma_1, \sigma_2, \ldots, \sigma_{i-1}, \sigma^+_i, \sigma_{i+1}, \ldots, \sigma_n\rangle$ is an optimal solution, which is in generalized weighted SPT-order.
		Applying the argument iteratively, we obtain the lemma.
		
		It remains to argue why the claim indeed holds.
		Let $B_\text{greedy}$ be the block that is the destination of $\sigma^+_i$ and let $j$ be the job in front of $j_i$ in $\langle\sigma_1, \ldots, \sigma_i\rangle$.
		Consider the case $i' = i+1$.
		By the assumption that $\pi$ is in generalized weighted SPT-order, $\sigma_{i'}$ can move $j_{i'}$ to a block between $B_\text{greedy}$ (inclusive) and $j$ only if at the respective position $j_{i}$ cannot be scheduled. 
		This, however, cannot be true due to the definition of $B_\text{greedy}$.
		Therefore, the improvement of $\sigma_{i'}$ is the same independent of whether it is applied in $\langle\sigma_1, \ldots, \sigma_i, \ldots \sigma_{i'-1}\rangle$ or $\langle\sigma_1, \ldots, \sigma^+_i, \ldots \sigma_{i'-1}\rangle$.
		The claim follows by inductively applying the argument to all $i' > i+1$.
	\end{proof}

	By the previous lemma, we have the final theorem of this section.
	
	\begin{theorem}
		The local search algorithm computes optimal solutions for the one-time setup problem in time $O(n\cdot log(n) \cdot K!)$.
		In combination with the \textsc{Transform} algorithm from \cref{sec:oneTime}, this yields an approximation algorithm with approximation factor $1+\sqrt{2}$ for our original problem.
	\end{theorem}
	
	\section{Arbitrary Number of Families}
	\label{sec:variableFamilies}
	In the previous section, we have seen that \modelMainW can be solved in time $O(n\cdot log(n) \cdot K!)$, which is polynomial for a fixed number $K$ of families.
	At this point, one might ask whether there are approximation algorithms running in time $\text{poly}(n, K)$, and whether the non-polynomial dependence on $K$ is inherent to \modelOS.
	The latter is indeed true because Woeginger has shown in his paper \cite{woeginger2003approximability} that different special cases  of the \modelDefPrec model, including one being equivalent to our (glued) \modelOS with the restriction that all job weights are $1$, are equally hard to approximate.
	Therefore, optimally solving the one-time setup problem is indeed NP-hard for non-constant $K$.
	On the positive side, we show how \modelMainW problems can be approximated in time $\text{poly}(n, K)$ in \cref{sec:lpApprox}.
	This approach, however, worsens the approximation by a factor of $2$ from $(1+\sqrt{2})$ to $2(1+\sqrt{2})$.
	Lastly we show that \modelMainW is inapproximable within factor $2-\varepsilon$, assuming a version of the Unique Games Conjecture, by applying results from Woeginger \cite{woeginger2003approximability} and Bansal and Khot \cite{bansal2009optimal}.
	
	\label{sec:reduction}
	
	\subsection{Approximation Algorithm}
	\label{sec:lpApprox}
	The general idea of our approximation algorithm is the same as for the case of a constant $K$: 
	We first solve \modelOS and then use \textsc{Transform} from \cref{sub:transform} to obtain a feasible schedule for our original \modelMainW problem.
	Recall that \modelOS is a special case of \modelDefPrec.
	As this problem has been studied a lot, there are different approximation algorithms in the literature and, for example, \cite{hall,motwani} provide $2$-approximation algorithms.
	Therefore, we conclude with the following theorem.
	
	\begin{theorem}
		\modelMainW can be approximated with an approximation factor of $2(1+\sqrt{2})$ in polynomial time.
	\end{theorem}
	
	\subsection{Lower Bound on the Approximability}
	\label{sec:lbApprox}
	\begin{theorem}
		Assuming a stronger version of the Unique Games Conjecture \cite{bansal2009optimal}, \modelMainW is inapproximable within $2-\varepsilon$ for any $\varepsilon > 0$.
	\end{theorem}
	\begin{proof}
		Woeginger \cite{woeginger2003approximability} showed that the general \modelDefPrec and some special cases of the problem have the same approximability threshold.
		Bansal and Khot \cite{bansal2009optimal} could prove that, assuming a stronger version of the Unique Games Conjecture, \modelDefPrec, and therefore also the special cases in \cite{woeginger2003approximability}, are inapproximable within $2-\varepsilon$ for any $\varepsilon > 0$.
		The special case we are interested in was defined by Woeginger as: \emph{[the] special case where every job has either $p_j=0$ and $w_j=1$, or $p_j=1$ and $w_j=0$, and where the existence of a precedence constraint $J_i\to J_j$ implies that $p_i=1$ and $w_i=0$, and that $p_j=0$ and $w_j=1$.}\cite{woeginger2003approximability} 
		It is easy to see that an $\alpha$-approximation for \modelMainW also yields an $\alpha$-approximation for the stated special case by transforming an instance of the special case in the following way:
		For every job $i$ with $w_i=0$ add a family $f_i$ with $s(f_i)=1$. 
		For every job $l$ with $w_l=1$ add a job $j_l$ with $w_{j_l}=1$. 
		For every precedence $J_i\to J_l$ add an operation $o_{i,l}$ to $j_l$ with $f(o_{i,l})=f_i$ and $p(o_{i,l})=0$. 
		It is easy to see that the optimal solutions of both problems have the same weight. 
		In both representations the difficult part is to decide the order of weight $0$ jobs or setups, respectively.
		All jobs or operations with processing time $0$ can be scheduled as early as possible in an optimal solution.
		
		Therefore we can conclude that \modelMainW has at least an equally high appoximability threshold as \modelDefPrec.
	\end{proof}
	
	\section{Simulation Results}
	\label{sec:simulation}
	To conclude our study of the problem, we also performed a simulation-based analysis of our approach to complement the theoretical results.
	To this end, we have taken a look at the conjecture from \cref{sub:transform} on the approximation quality of our algorithm for a constant number $K$ of families.
	Additionally, we propose a way to improve its performance on randomly created instances.
	
	\subsection{Approximation Quality for Constant $K$}
	\label{sec:simApprox}
\begin{figure}[p]
		\centering
		\begin{subfigure}[b]{0.45\textwidth}
			\centering
			\def\svgwidth{\linewidth}
			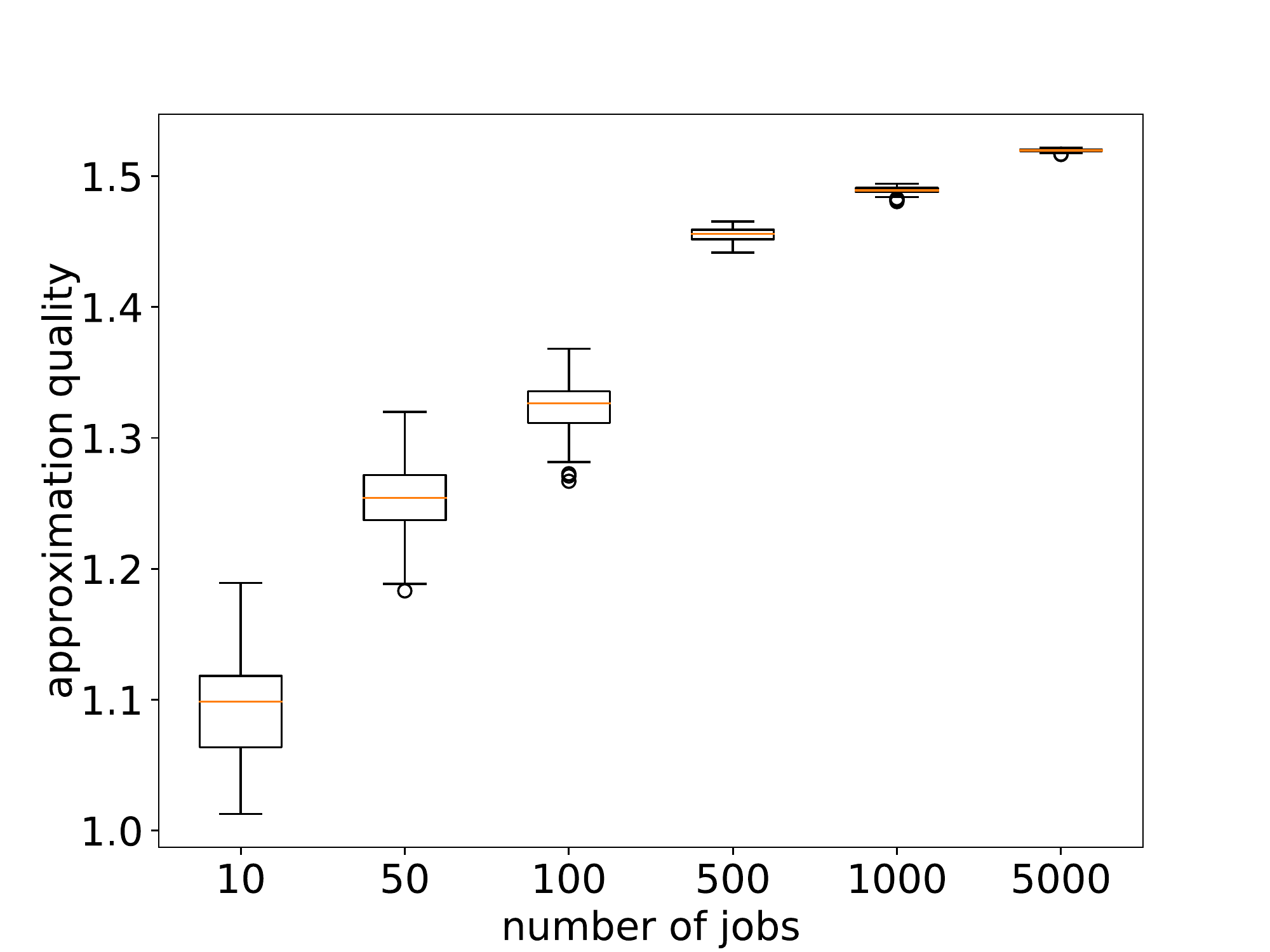
			\caption{}
		\end{subfigure}
		\begin{subfigure}[b]{0.45\textwidth}  
			\centering 
			\def\svgwidth{\linewidth}
			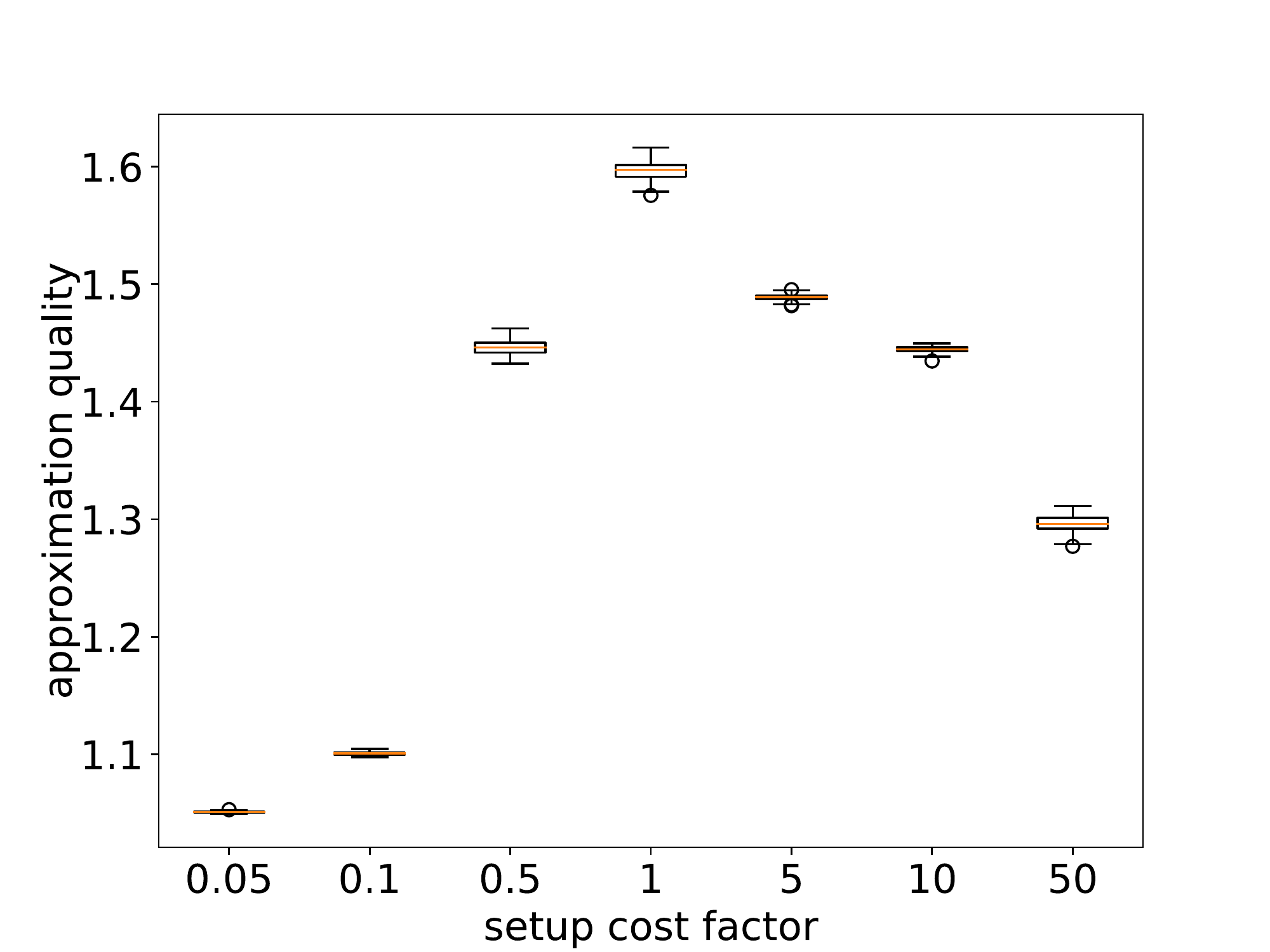
			\caption{}
		\end{subfigure}
		\vskip\baselineskip
		\begin{subfigure}[b]{0.45\textwidth}   
			\centering 
			\def\svgwidth{\linewidth}
			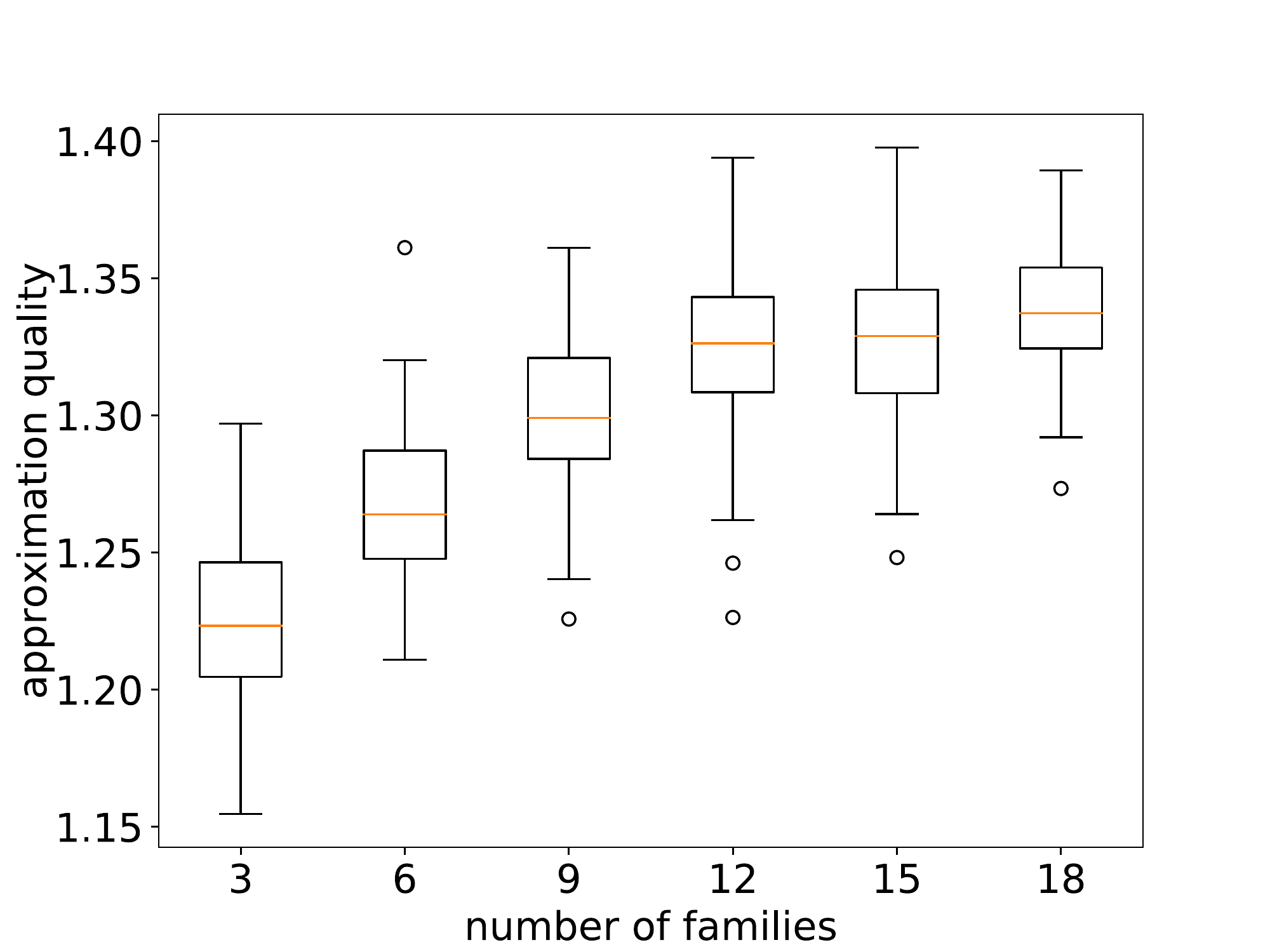
			\caption{} \label{fig:families}
		\end{subfigure}
		\begin{subfigure}[b]{0.45\textwidth}   
			\centering 
			\def\svgwidth{\linewidth}
			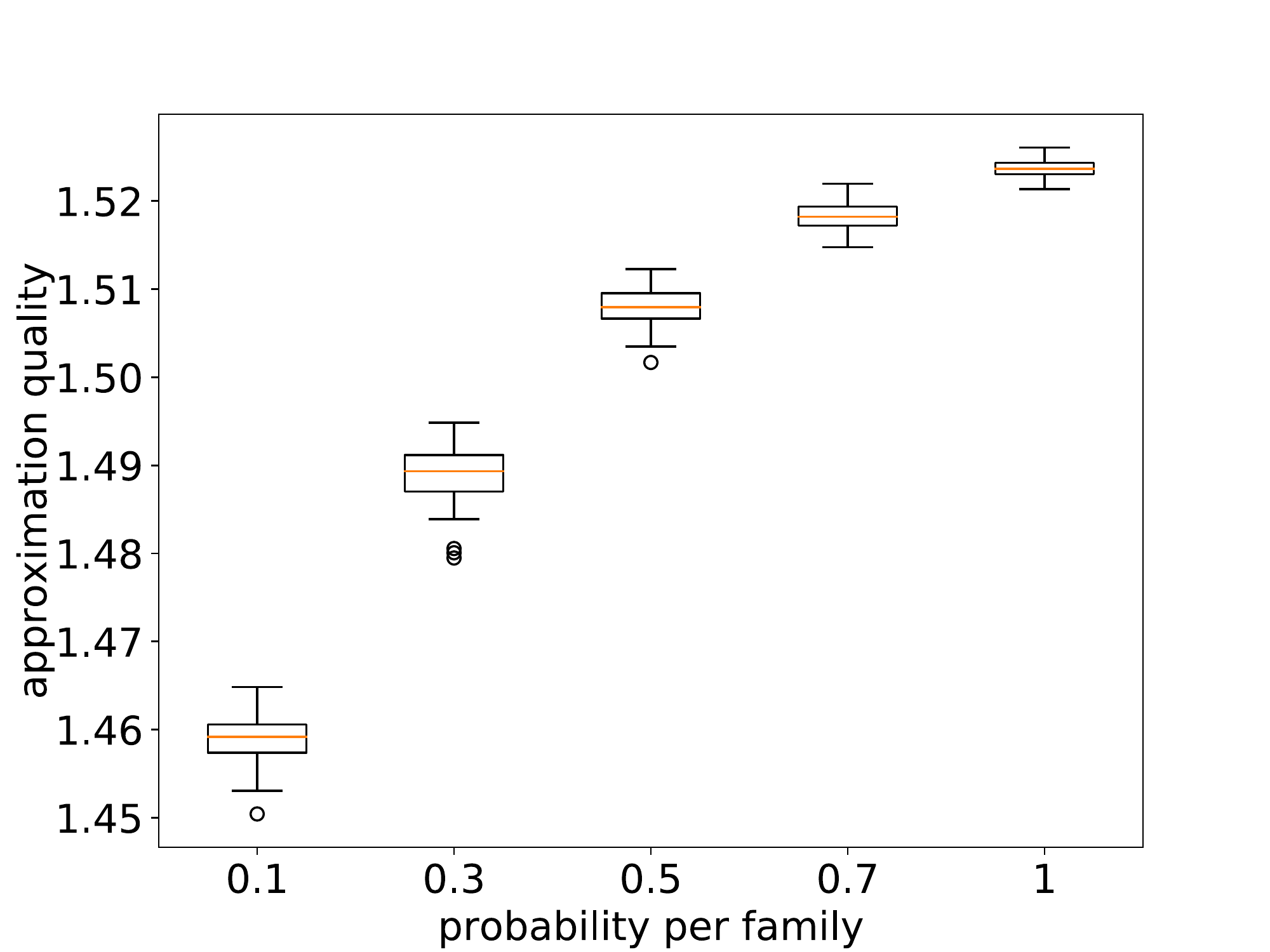
			\caption{}
		\end{subfigure}
		\caption{Approximation quality observed in simulations for (unweighted) total completion time depending on different parameters.
			Unless stated otherwise, the number of jobs is $1000$, the setup cost factor is $5$, the number of families is $5$ and the probability per familie is set to $0.3$.
			In \cref{fig:families} the number of jobs is $50$.} 
		\label{fig:sim_approx}
	\end{figure}

	In \cref{sub:transform} we conjectured that on instances less artificial than the instances constructed to show the tightness of the analysis of \textsc{Transform}, the approximation factor of our approach should rather be upper bounded by $1+\frac{2}{\pullfactor}$ than by $1+\pullfactor$ for moderate values of \pullfactor.
	To give evidence for this conjecture, we simulated our approach on randomly created instances with small constant pull factors \pullfactor.
	We evaluated the \emph{approximation quality} given by the ratio of the total completion time and a lower bound on \opt given by the solution to the one-time setup problem.
	The randomly created instances are based on processing times for operations that are randomly drawn from a normal distribution.
	However, for other distributions like log-normal, uniform, and Weibull we obtained very similar results.
	The setup costs of each family were set to the average processing time of that family's operations multiplied by a \emph{setup cost factor}, which we varied in different experiments. 
	Similarly, each job contains an operation of a family with probability \emph{probability per family}, also varied in different experiments.
	\Cref{fig:sim_approx} shows the typical behavior we observed in our simulations, here exemplarily for the case of a pull factor $\pullfactor = 2$.
	As one can see, the observed approximation quality always stayed below $2$ and never came close to the theoretically possible $3$ in our simulations.
	
	\newpage
	\subsection{Heuristical Improvements}
	\label{sec:simPullFactor}
\begin{figure}
		\centering
		\def\svgwidth{0.45\linewidth}
		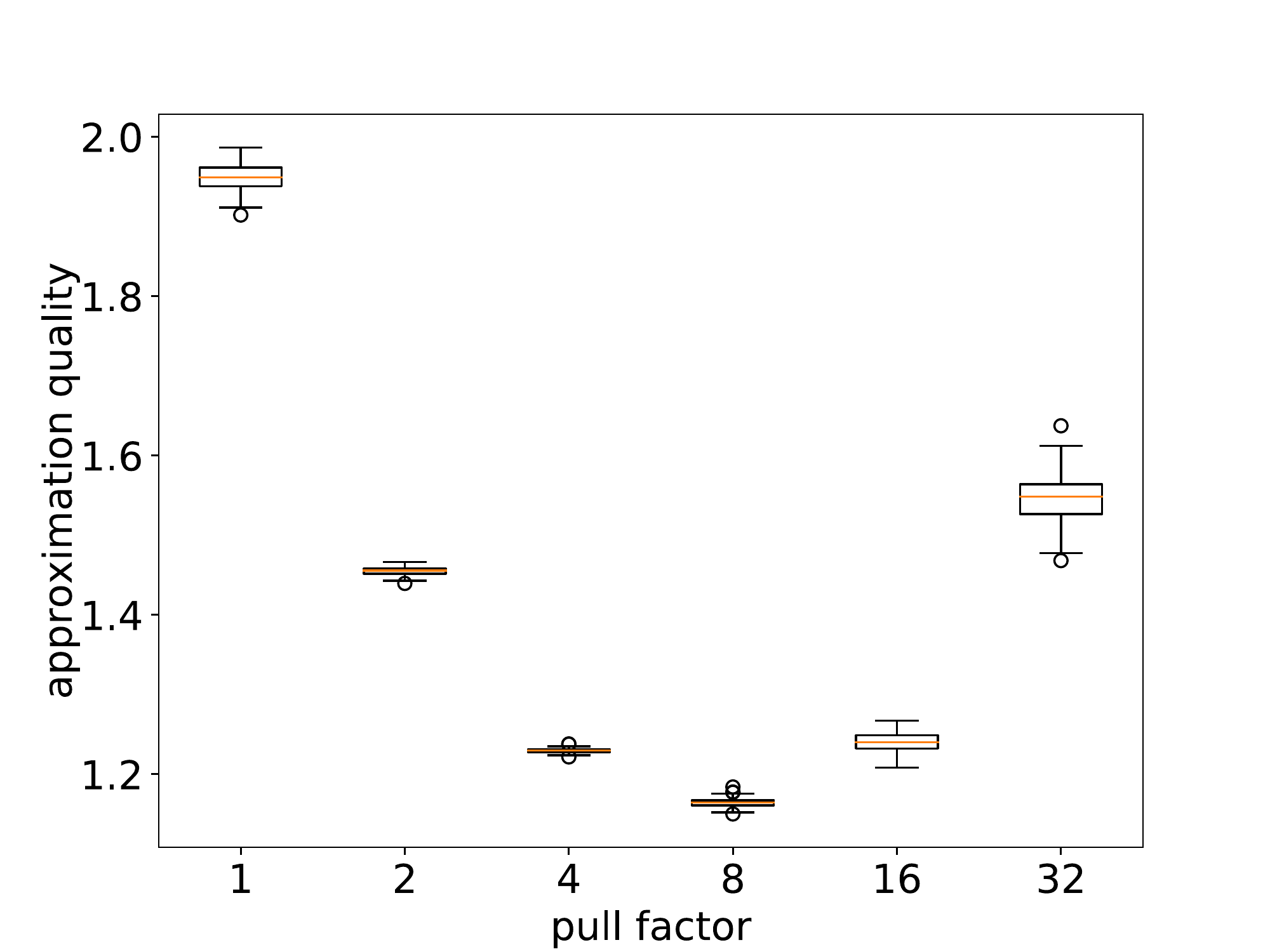
		\caption{Approximation quality observed in simulations for (unweighted) total completion time depending on the pull factor \pullfactor.
			The number of jobs is set to $500$, the setup cost factor is $5$, the number of families is $5$ and the probability per family is $0.3$.}
		\label{fig:sim_pf}
	\end{figure}

	In our \textsc{Transform} algorithm, we build batches given a solution to the one-time setup problem by moving operations to the front until a batch becomes ``sufficiently large'' to justify a setup for the respective family.
	The term ``sufficiently large'' is thereby determined based on the pull factor \pullfactor.
	Our analysis of \textsc{Transform} was shown to be tight and theoretically the best \pullfactor is $\sqrt{2}$.
	However, we also already conjectured in \cref{sub:transform} that practically other values for \pullfactor might lead to superior performance. 
	Therefore, this parameter gives a natural option to tune the algorithm.
	One might expect that the best pull factor depends on various parameters of an instance such as the processing times or number of operations per family.
	\Cref{fig:sim_pf} shows a typical result of our simulations proving that there indeed is much room for improvement compared to a pull factor of $\sqrt{2}$ if randomly created instances are considered.
	
	\section{Future Work}
	For future work it might be interesting whether there is a better algorithm for transforming solutions for the one-time setup problem to their respective original problem.
	One could also try to improve the approximation factor by designing algorithms that directly solve our original problem without the detour via the one-time setup problem.
	Another interesting direction for the future is the question whether our lower bound can be increased.
	For the special case with a constant number of families, the question whether that problem is already NP-hard also remains open.
	
	\section*{Acknowledgments}
	We thank the anonymous reviewers who helped us improve the quality of this paper with useful comments and pointing us towards important reference material.
	
	%
	%
	%
	\bibliographystyle{splncs04}
	\bibliography{references}
	
\end{document}

%% file: tightness.pdf_tex
\begingroup%
  \makeatletter%
  \providecommand\color[2][]{%
    \errmessage{(Inkscape) Color is used for the text in Inkscape, but the package 'color.sty' is not loaded}%
    \renewcommand\color[2][]{}%
  }%
  \providecommand\transparent[1]{%
    \errmessage{(Inkscape) Transparency is used (non-zero) for the text in Inkscape, but the package 'transparent.sty' is not loaded}%
    \renewcommand\transparent[1]{}%
  }%
  \providecommand\rotatebox[2]{#2}%
  \newcommand*\fsize{\dimexpr\f@size pt\relax}%
  \newcommand*\lineheight[1]{\fontsize{\fsize}{#1\fsize}\selectfont}%
  \ifx\svgwidth\undefined%
    \setlength{\unitlength}{308.66763618bp}%
    \ifx\svgscale\undefined%
      \relax%
    \else%
      \setlength{\unitlength}{\unitlength * \real{\svgscale}}%
    \fi%
  \else%
    \setlength{\unitlength}{\svgwidth}%
  \fi%
  \global\let\svgwidth\undefined%
  \global\let\svgscale\undefined%
  \makeatother%
  \begin{picture}(1,0.09190063)%
    \lineheight{1}%
    \setlength\tabcolsep{0pt}%
    \put(0,0){\includegraphics[width=\unitlength,page=1]{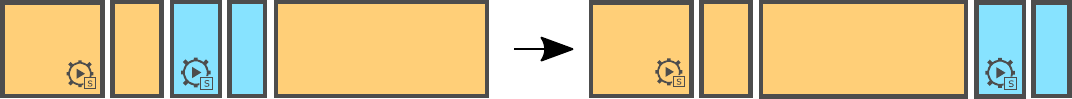}}%
    \put(0.03712869,0.05306389){\color[rgb]{0,0,0}\makebox(0,0)[lt]{\lineheight{1.25}\smash{\begin{tabular}[t]{l}$1$\end{tabular}}}}%
    \put(0.11714406,0.0533075){\color[rgb]{0,0,0}\makebox(0,0)[lt]{\lineheight{1.25}\smash{\begin{tabular}[t]{l}$\varepsilon$\end{tabular}}}}%
    \put(0.17208436,0.0533075){\color[rgb]{0,0,0}\makebox(0,0)[lt]{\lineheight{1.25}\smash{\begin{tabular}[t]{l}$\varepsilon$\end{tabular}}}}%
    \put(0.21994911,0.0533075){\color[rgb]{0,0,0}\makebox(0,0)[lt]{\lineheight{1.25}\smash{\begin{tabular}[t]{l}$\varepsilon$\end{tabular}}}}%
    \put(0.31463351,0.0532814){\color[rgb]{0,0,0}\makebox(0,0)[lt]{\lineheight{1.25}\smash{\begin{tabular}[t]{l}$\pullfactor-2\varepsilon$\end{tabular}}}}%
    \put(0.58660993,0.05478402){\color[rgb]{0,0,0}\makebox(0,0)[lt]{\lineheight{1.25}\smash{\begin{tabular}[t]{l}$1$\end{tabular}}}}%
    \put(0.66662487,0.05502764){\color[rgb]{0,0,0}\makebox(0,0)[lt]{\lineheight{1.25}\smash{\begin{tabular}[t]{l}$\varepsilon$\end{tabular}}}}%
    \put(0.92250244,0.05302019){\color[rgb]{0,0,0}\makebox(0,0)[lt]{\lineheight{1.25}\smash{\begin{tabular}[t]{l}$\varepsilon$\end{tabular}}}}%
    \put(0.97036711,0.05302019){\color[rgb]{0,0,0}\makebox(0,0)[lt]{\lineheight{1.25}\smash{\begin{tabular}[t]{l}$\varepsilon$\end{tabular}}}}%
    \put(0.7640626,0.05390382){\color[rgb]{0,0,0}\makebox(0,0)[lt]{\lineheight{1.25}\smash{\begin{tabular}[t]{l}$\pullfactor-2\varepsilon$\end{tabular}}}}%
  \end{picture}%
\endgroup%

%% file: graph_x_for_different_number_of_jobs_normal.pdf_tex
\begingroup%
  \makeatletter%
  \providecommand\color[2][]{%
    \errmessage{(Inkscape) Color is used for the text in Inkscape, but the package 'color.sty' is not loaded}%
    \renewcommand\color[2][]{}%
  }%
  \providecommand\transparent[1]{%
    \errmessage{(Inkscape) Transparency is used (non-zero) for the text in Inkscape, but the package 'transparent.sty' is not loaded}%
    \renewcommand\transparent[1]{}%
  }%
  \providecommand\rotatebox[2]{#2}%
  \newcommand*\fsize{\dimexpr\f@size pt\relax}%
  \newcommand*\lineheight[1]{\fontsize{\fsize}{#1\fsize}\selectfont}%
  \ifx\svgwidth\undefined%
    \setlength{\unitlength}{576bp}%
    \ifx\svgscale\undefined%
      \relax%
    \else%
      \setlength{\unitlength}{\unitlength * \real{\svgscale}}%
    \fi%
  \else%
    \setlength{\unitlength}{\svgwidth}%
  \fi%
  \global\let\svgwidth\undefined%
  \global\let\svgscale\undefined%
  \makeatother%
  \begin{picture}(1,0.75)%
    \lineheight{1}%
    \setlength\tabcolsep{0pt}%
    \put(0,0){\includegraphics[trim=0 0 55 45, clip, width=\unitlength,page=1]{graph_x_for_different_number_of_jobs_normal.pdf}}%
  \end{picture}%
\endgroup%

%% file: graph_x_for_different_setups_normal.pdf_tex
\begingroup%
  \makeatletter%
  \providecommand\color[2][]{%
    \errmessage{(Inkscape) Color is used for the text in Inkscape, but the package 'color.sty' is not loaded}%
    \renewcommand\color[2][]{}%
  }%
  \providecommand\transparent[1]{%
    \errmessage{(Inkscape) Transparency is used (non-zero) for the text in Inkscape, but the package 'transparent.sty' is not loaded}%
    \renewcommand\transparent[1]{}%
  }%
  \providecommand\rotatebox[2]{#2}%
  \newcommand*\fsize{\dimexpr\f@size pt\relax}%
  \newcommand*\lineheight[1]{\fontsize{\fsize}{#1\fsize}\selectfont}%
  \ifx\svgwidth\undefined%
    \setlength{\unitlength}{576bp}%
    \ifx\svgscale\undefined%
      \relax%
    \else%
      \setlength{\unitlength}{\unitlength * \real{\svgscale}}%
    \fi%
  \else%
    \setlength{\unitlength}{\svgwidth}%
  \fi%
  \global\let\svgwidth\undefined%
  \global\let\svgscale\undefined%
  \makeatother%
  \begin{picture}(1,0.75)%
    \lineheight{1}%
    \setlength\tabcolsep{0pt}%
    \put(0,0){\includegraphics[trim=0 0 55 45, clip, width=\unitlength,page=1]{graph_x_for_different_setups_normal.pdf}}%
  \end{picture}%
\endgroup%

%% file: graph_x_for_different_number_families_normal.pdf_tex
\begingroup%
  \makeatletter%
  \providecommand\color[2][]{%
    \errmessage{(Inkscape) Color is used for the text in Inkscape, but the package 'color.sty' is not loaded}%
    \renewcommand\color[2][]{}%
  }%
  \providecommand\transparent[1]{%
    \errmessage{(Inkscape) Transparency is used (non-zero) for the text in Inkscape, but the package 'transparent.sty' is not loaded}%
    \renewcommand\transparent[1]{}%
  }%
  \providecommand\rotatebox[2]{#2}%
  \newcommand*\fsize{\dimexpr\f@size pt\relax}%
  \newcommand*\lineheight[1]{\fontsize{\fsize}{#1\fsize}\selectfont}%
  \ifx\svgwidth\undefined%
    \setlength{\unitlength}{576bp}%
    \ifx\svgscale\undefined%
      \relax%
    \else%
      \setlength{\unitlength}{\unitlength * \real{\svgscale}}%
    \fi%
  \else%
    \setlength{\unitlength}{\svgwidth}%
  \fi%
  \global\let\svgwidth\undefined%
  \global\let\svgscale\undefined%
  \makeatother%
  \begin{picture}(1,0.75)%
    \lineheight{1}%
    \setlength\tabcolsep{0pt}%
    \put(0,0){\includegraphics[trim=0 0 55 45, clip, width=\unitlength,page=1]{graph_x_for_different_number_families_normal.pdf}}%
  \end{picture}%
\endgroup%

%% file: graph_x_for_different_fam_prob_normal.pdf_tex
\begingroup%
  \makeatletter%
  \providecommand\color[2][]{%
    \errmessage{(Inkscape) Color is used for the text in Inkscape, but the package 'color.sty' is not loaded}%
    \renewcommand\color[2][]{}%
  }%
  \providecommand\transparent[1]{%
    \errmessage{(Inkscape) Transparency is used (non-zero) for the text in Inkscape, but the package 'transparent.sty' is not loaded}%
    \renewcommand\transparent[1]{}%
  }%
  \providecommand\rotatebox[2]{#2}%
  \newcommand*\fsize{\dimexpr\f@size pt\relax}%
  \newcommand*\lineheight[1]{\fontsize{\fsize}{#1\fsize}\selectfont}%
  \ifx\svgwidth\undefined%
    \setlength{\unitlength}{576bp}%
    \ifx\svgscale\undefined%
      \relax%
    \else%
      \setlength{\unitlength}{\unitlength * \real{\svgscale}}%
    \fi%
  \else%
    \setlength{\unitlength}{\svgwidth}%
  \fi%
  \global\let\svgwidth\undefined%
  \global\let\svgscale\undefined%
  \makeatother%
  \begin{picture}(1,0.75)%
    \lineheight{1}%
    \setlength\tabcolsep{0pt}%
    \put(0,0){\includegraphics[trim=0 0 55 45, clip, width=\unitlength,page=1]{graph_x_for_different_fam_prob_normal.pdf}}%
  \end{picture}%
\endgroup%

%% file: graph_x_for_different_pull_factors_normal.pdf_tex
\begingroup%
  \makeatletter%
  \providecommand\color[2][]{%
    \errmessage{(Inkscape) Color is used for the text in Inkscape, but the package 'color.sty' is not loaded}%
    \renewcommand\color[2][]{}%
  }%
  \providecommand\transparent[1]{%
    \errmessage{(Inkscape) Transparency is used (non-zero) for the text in Inkscape, but the package 'transparent.sty' is not loaded}%
    \renewcommand\transparent[1]{}%
  }%
  \providecommand\rotatebox[2]{#2}%
  \newcommand*\fsize{\dimexpr\f@size pt\relax}%
  \newcommand*\lineheight[1]{\fontsize{\fsize}{#1\fsize}\selectfont}%
  \ifx\svgwidth\undefined%
    \setlength{\unitlength}{576bp}%
    \ifx\svgscale\undefined%
      \relax%
    \else%
      \setlength{\unitlength}{\unitlength * \real{\svgscale}}%
    \fi%
  \else%
    \setlength{\unitlength}{\svgwidth}%
  \fi%
  \global\let\svgwidth\undefined%
  \global\let\svgscale\undefined%
  \makeatother%
  \begin{picture}(1,0.75)%
    \lineheight{1}%
    \setlength\tabcolsep{0pt}%
    \put(0,0){\includegraphics[trim=0 0 55 45, clip, width=\unitlength,page=1]{graph_x_for_different_pull_factors_normal.pdf}}%
  \end{picture}%
\endgroup%

%% file: orderSchedulingSetups.bbl
\begin{thebibliography}{10}
\providecommand{\url}[1]{\texttt{#1}}
\providecommand{\urlprefix}{URL }
\providecommand{\doi}[1]{https://doi.org/#1}

\bibitem{ali3}
Allahverdi, A.: The third comprehensive survey on scheduling problems with
  setup times/costs. European Journal of Operational Research  \textbf{246}(2),
   345--378 (2015)

\bibitem{ali1}
Allahverdi, A., Gupta, J.N., Aldowaisan, T.: A review of scheduling research
  involving setup considerations. Omega  \textbf{27}(2),  219--239 (1999)

\bibitem{ali2}
Allahverdi, A., Ng, C.T., Cheng, T.C.E., Kovalyov, M.Y.: A survey of scheduling
  problems with setup times or costs. European Journal of Operational Research
  \textbf{187}(3),  985--1032 (2008)

\bibitem{bansal2009optimal}
Bansal, N., Khot, S.: Optimal long code test with one free bit. In: Proceedings
  of the 50th Annual IEEE Symposium on Foundations of Computer Science (FOCS).
  pp. 453--462. IEEE (2009)

\bibitem{motwani}
Chekuri, C., Motwani, R.: Precedence constrained scheduling to minimize sum of
  weighted completion times on a single machine. Discrete Applied Mathematics
  \textbf{98}(1-2),  29--38 (1999)

\bibitem{correa15}
Correa, J.R., Marchetti{-}Spaccamela, A., Matuschke, J., Stougie, L., Svensson,
  O., Verdugo, V., Verschae, J.: Strong {LP} formulations for scheduling
  splittable jobs on unrelated machines. Mathematical Programming
  \textbf{154}(1-2),  305--328 (2015)

\bibitem{splittingVsSetups}
Correa, J.R., Verdugo, V., Verschae, J.: Splitting versus setup trade-offs for
  scheduling to minimize weighted completion time. Operations Research Letters
  \textbf{44}(4),  469--473 (2016)

\bibitem{saksCompletionTime}
Divakaran, S., Saks, M.E.: Approximation algorithms for problems in scheduling
  with set-ups. Discrete Applied Mathematics  \textbf{156}(5),  719--729 (2008)

\bibitem{agreeable}
Gerodimos, A.E., Glass, C.A., Potts, C.N., Tautenhahn, T.: Scheduling
  multi-operation jobs on a single machine. Annals {OR}  \textbf{92},  87--105
  (1999)

\bibitem{hall}
Hall, L.A., Schulz, A.S., Shmoys, D.B., Wein, J.: Scheduling to minimize
  average completion time: Off-line and on-line approximation algorithms.
  Mathematics of Operations Research  \textbf{22}(3),  513--544 (1997)

\bibitem{sfb901}
Happe, M., {Meyer auf der Heide}, F., Kling, P., Platzner, M., Plessl, C.:
  {On-The-Fly Computing: A Novel Paradigm for Individualized {IT} Services}.
  In: Proceedings of the 16th {IEEE} International Symposium on
  Object/Component/Service-Oriented Real-Time Distributed Computing (ISORC).
  pp. 1--10. {IEEE} Computer Society (2013)

\bibitem{martenConfigIP}
Jansen, K., Klein, K., Maack, M., Rau, M.: Empowering the configuration-ip -
  new {PTAS} results for scheduling with setups times. In: Proceedings of the
  10th Innovations in Theoretical Computer Science Conference (ITCS). LIPIcs,
  vol.~124, pp. 44:1--44:19. Schloss Dagstuhl - Leibniz-Zentrum fuer Informatik
  (2019)

\bibitem{kiel}
Jansen, K., Maack, M., M{\"{a}}cker, A.: Scheduling on (un-)related machines
  with setup times. In: Proceedings of the 2019 {IEEE} International Parallel
  and Distributed Processing Symposium (IPDPS). pp. 145--154

\bibitem{lawler1978sequencing}
Lawler, E.L.: Sequencing jobs to minimize total weighted completion time
  subject to precedence constraints. In: Annals of Discrete Mathematics,
  vol.~2, pp. 75--90. Elsevier (1978)

\bibitem{lenstra1978complexity}
Lenstra, J.K., Kan, A.H.G.R.: Complexity of scheduling under precedence
  constraints. Operations Research  \textbf{26}(1),  22--35 (1978)

\bibitem{orderSurvey}
Leung, J.Y., Li, H., Pinedo, M.: Order scheduling models: an overview. In:
  Multidisciplinary scheduling: theory and applications, pp. 37--53. Springer
  (2005)

\bibitem{monma1989complexity}
Monma, C.L., Potts, C.N.: On the complexity of scheduling with batch setup
  times. Operations research  \textbf{37}(5),  798--804 (1989)

\bibitem{hardness}
Ng, C.T., Cheng, T.C.E., Yuan, J.J.: Strong {NP}-hardness of the single machine
  multi-operation jobs total completion time scheduling problem. Information
  Processing Letters  \textbf{82}(4),  187--191 (2002)

\bibitem{wspt}
Smith, W.E.: Various optimizers for single-stage production. Naval Research
  Logistics Quarterly  \textbf{3}(1-2),  59--66 (1956)

\bibitem{woeginger2003approximability}
Woeginger, G.J.: On the approximability of average completion time scheduling
  under precedence constraints. Discrete Applied Mathematics  \textbf{131}(1),
  237--252 (2003)

\end{thebibliography}
